\documentstyle[12pt]{article}
\begin{document}
\begin{center}
{\Large \bf On $p$-adic path integral }

\bigskip
{\large   Branko  Dragovich }

\date{}

\smallskip

{\it Institute of Physics, P.O.Box 57, 11001 Belgrade, Yugoslavia}

\end{center}

\noindent{Dedicated to the memory}

\noindent{of N.N.Bogolyubov}

\begin{abstract} 
Feynman's path integral is generalized to quantum mechanics on
$p$-adic space and time. Such $p$-adic path integral is analytically
evaluated for quadratic Lagrangians. Obtained result has the same 
form as that one in ordinary quantum mechanics.
\end{abstract} 


{\bf 1.} It is well known that dynamical evolution of any one-dimensional
quantum-mechanical system, described by a wave function $\Psi (x,t)$, 
is given by
$$
   \Psi (x'',t'') = \int {\cal K}(x'',t'';x',t')\Psi(x',t')dx' , \eqno(1)
$$
where ${\cal K}(x'',t'';x',t')$  is the kernel of the corresponding unitary operator 
acting as follows:
$$
   \Psi (t'') = {\cal U}(t'',t')\Psi (t') .              \eqno(2)
$$

$ {\cal K}(x'',t'';x',t')$ is also called Green's function, or the quantum-mechanical
propagator, and the probability amplitude to go a particle from a
point $ (x',t')$ to a point $(x'',t'')$. One can easily deduce
the following three general properties:
$$
  \int{\cal K}(x'',t'';x,t) {\cal K}(x,t;x',t') dx = {\cal K}(x'',t'';x't') ,      \eqno(3)
$$
$$
   \int {\cal K}^{*}(x'',t'';x',t') {\cal K}(x'',t'';x,t') dx'' = \delta (x'-x) ,       \eqno(4)
$$
$$
   {\cal K}(x'',t'';x',t'') = lim_{t'\to t''} {\cal K}(x'',t'';x',t') = \delta (x''-x') .    \eqno(5)
$$
Since all information on quantum dynamics can be deduced from
the propagator ${\cal K}(x'',t'';x',t')$ it can be regarded as the basic ingredient of quantum
theory. In Feynman's formulation \cite{feynman} of quantum mechanics,
${\cal K}(x'',t'';x',t')$ was postulated to be the path integral
$$
   {\cal K}(x'',t'';x',t') = \int \exp{\left(\frac{2\pi i}{h} \int_{t'}^{t''} 
   L(\dot{q},q,t)dt\right)} {\cal D}q ,  \eqno(6)
$$
where $x'' = q(t'')$ and $x' = q(t')$, and $h$ is the Planck constant.

In its original form, the path integral (6) is the limit of 
the corresponding multiple integral of $n-1$ variables
$q_i = q(t_i), \ \   (i=1,2,...,n-1),$ when $n\to\infty$.
For the half of century of its history, the path integral
has been a subject of permanent interest in theoretical
and mathematical physics. At present days (see, e.g. \cite{firenca})
it is one of the most profound and promising approaches to 
foundations of quantum theory (in particular, quantum field theory
and superstring theory). Feynman's path integral is inevitable
in formulation of $p$-adic \cite{vlad-vol} and adelic \cite{branko1}
quantum mechanics. It is worth
noting that just Feynman's path integral approach enables natural
foundation of quantum theory on $p$-adic and adelic spaces.

\smallskip

{\bf 2.} Recall that the set of rational numbers $Q$ plays an
important role in mathematics as well as in physics. From algebraic 
point of view, $Q$ is the simplest number field of characteristic
$0$. The usual absolute value and $p$-adic valuation ($p$ is any 
of prime numbers) exhaust all possible non-trivial norms on
$Q$ \cite{schik}. Completion of $Q$ with respect to metrics 
induced by these norms leads to the field of real numbers $R$ and
the fields of $p$-adic numbers $Q_p$, $(p=2,3,5,...)$.
Thus $Q$ is dense in $R$ and all $Q_p$.
From physical point of view, all numerical results of measurements
are rational numbers. However, theoretical models  of physical 
systems are traditionally constructed using real and complex numbers.
One can ask the following question: Why real (and complex) numbers
are so good in description of usual physical phenomena, and, is there
any aspect of physical reality which has to be described by $p$-adic
numbers. Construction of $p$-adic models and their appropriate interpretation
can gradually give answer to this question.
Since 1987, there have been many publications (for a review,
see, e.g. \cite{freund,vvz,andrei} ) on possible applications of 
$p$-adic numbers in  modern theoretical and mathematical physics.
For a systematic approach to this subject, $p$-adic \cite{vlad-vol}
and adelic \cite{branko1} quantum mechanics have been formulated.

Recall also that any $p$-adic number $x\in Q_p$ can be presented
as the following infinite expansion
$$
  x = p^\nu (x_0 + x_1p + x_2p^2 + \cdots ) , \ \ \nu\in Z ,
$$
where $x_i = 0,1,...,p-1$ are digits.
We will use the Gauss integral \cite{vvz}
$$
   \int_{Q_p} \chi_p (\alpha x^2 + \beta x)dx =
   \lambda_p(\alpha) \mid 2\alpha \mid_p^{-\frac{1}{2}}
   \chi_p\left( - \frac{\beta^2}{4\alpha}\right) , \ \ \alpha\neq 0 ,    
$$
where $\chi_p (a) = \exp (2\pi i \{ a\}_p)$ is the additive character,
and $\{ a\}_p$ is the fractional part of $a\in Q_p$. $\lambda_p(x)$
is a complex-valued arithmetic function (for a definition, see \cite{vvz}) 
with the following properties:
$$
   \lambda_p(0) = 1 ,\ \lambda_p(a^2x) = \lambda_p(x) , 
   \ \lambda_p(x)\lambda_p(y) = \lambda_p(x+y)\lambda_p(x^{-1}+y^{-1}) ,
   \ \lambda_p^{*}(x)\lambda_p(x) = 1 .
$$

\smallskip

{\bf 3.} $p$-Adic quantum mechanics, we are interested in, contains
complex-valued functions of $p$-adic arguments. There is not the
corresponding Schr$\ddot{o}$dinger equation, but Feynman's path integral approach
seems to be quite natural. Feynman's path integral for $p$-adic
propagator ${\cal K}_p(x'',t'';x',t')$, where ${\cal K}_p$ is 
complex-valued and $x'',x',t'',t'$
are $p$-adic variables, is a direct $p$-adic generalization of (6),
i.e.
$$
   {\cal K}_p (x'',t'';x',t') = \int \chi_p \left( -\frac{1}{h} 
   \int_{t'}^{t''} L(\dot{q},q,t)dt\right) {\cal D}q ,    \eqno(7)
$$
where $ \chi_p(a)$ is $p$-adic additive character.
The Planck constant $h$ in (6) and (7) is the same rational number.
Integral $ \int_{t'}^{t''}L(\dot{q},q,t)dt$ we consider as 
the difference of antiderivative (without pseudoconstants) of
$L(\dot{q},q,t) $ in final $(t'')$ and initial $(t')$ times.
${\cal D}q = \prod_{i=1}^{n-1}dq(t_i)$, where $dq(t_i)$ is the $p$-adic 
Haar measure. Thus, $p$-adic 
path integral is the limit of the multiple Haar integral when
$n\to\infty$. To calculate (7) in this way  one has to introduce
some order on $t\in Q_p$, and it is successfully done in Ref.
\cite{zelenov}. On previous investigations of $p$-adic path
integral one can see \cite{branko2}. Our main task here  is 
derivation of the exact result for $p$-adic Feynman's path integral
(7) for the general case of Lagrangians
$L(\dot{q},q,t)$, which are quadratic polynomials in $\dot{q}$
and $q$, without making time discretization.

A general quadratic Lagrangian can be written as follows:
$$
   L(\dot{q},q,t) = \frac{1}{2}\frac{\partial^2 L_0}{\partial\dot{q}^2}\dot{q}^2
   + \frac{\partial L_0}{\partial\dot{q}} \dot{q}    +  
   \frac{\partial^2 L_0}{\partial\dot{q}\partial q}\dot{q}q + L_0
   +  \frac{\partial L_0}{\partial q} q  +  
   \frac{1}{2}\frac{\partial^2 L_0}{\partial q^2} q^2 ,
   \eqno(8)
$$
where index $0$ denotes that the Taylor expansion of $L(\dot{q},q,t)$
is around $\dot{q}=q=0$. The Euler-Lagrange equation of motion is
$$
 \frac{\partial^2 L_0}{\partial\dot{q}^2} \ddot{q} +
 \frac{d}{dt} \left( \frac{\partial^2 L_0}{\partial\dot{q}^2} \right) \dot{q} + 
 \left[ \frac{d}{dt} \left( \frac{\partial^2 L_0}{\partial\dot{q}\partial q}  \right) - 
 \frac{\partial^2 L_0}{\partial q^2}  \right] q = \frac{\partial L_0}{\partial q}
 - \frac{d}{dt}\left( \frac{\partial L_0}{\partial\dot{q}}  \right) .
 \eqno(9)
$$
General solution of (9) is
$$
q \equiv x(t) = C_1 x_1(t) + C_2 x_2(t) + w(t) ,   \eqno(10)
$$
where $x_1(t)$ and $x_2(t)$ are two linearly independent solutions
of the corresponding homogeneous equation, and $w(t)$ is a particular 
solution of the complete equation (9). Note that $x(t)$ denotes 
the classical trajectory. Imposing the boundary conditions
$x'=x(t')$ and    $x''=x(t'')$, constants of integration $C_1$ and  $C_2$
become:
$$
   C_1 \equiv C_1 (t'',t') = \frac{(x'-w')x''_2 - (x''-w'')x'_2}{x''_2x'_1 - x''_1x'_2},     \eqno(11a)
$$
$$
   C_2 \equiv C_2 (t'',t')  =  \frac{(x''-w'')x'_1 - (x'-w')x''_1}{x''_2x'_1 - x''_1x'_2}.   \eqno(11b)
$$
Since $C_1(t'',t')$ and   $C_2(t'',t')$ are linear in $x''$  and $x'$,
the corresponding classical action  $\bar{S}(x'',t'';x',t') = \int_{t'}^{t''} L(\dot{x},x,t)dt$ 
is quadratic in $x''$ and
$x'$. Note that the above expressions have the same form in $R$ and $Q_p$.

Quantum fluctuations lead to deviations of classical trajectory and
any quantum path may be presented as $q(t) = x(t) + y(t)$, where
$y'=y(t')=0$  and  $y''=y(t'')=0$.  The corresponding Taylor expansion
of $S[q]$ around classical path $x(t)$  is
$$
   S[q] = S[x+y] = S[x] + \frac{1}{2!} \delta^2 S[x] = S[x] + 
   \frac{1}{2}\int_{t'}^{t''} \left( \dot{y}\frac{\partial}{\partial \dot{q}} + 
   y\frac{\partial}{\partial q}   \right)^{(2)} L(\dot{q},q,t)dt ,  \eqno(12)
$$
where we used $\delta S[x] = 0$. We have now
$$
   {\cal K}_p (x'',t'';x',t') = \chi_p \left( -\frac{1}{h} S[x] \right) \int
   \chi_p \left(- \frac{1}{2h} \int_{t'}^{t''} \left( \dot{y}\frac{\partial}{\partial\dot{q}} +
    y \frac{\partial}{\partial q} \right)^{(2)}L(\dot{q},q,t)dt  \right) {\cal D}y 
    \eqno(13)
$$
with $ y''=y'=0 $ and $S[x] = \bar{S}(x'',t'';x',t')$.

Note that ${\cal K}_p(x'',t'';x',t')$  has the form
$$
    {\cal K}_p(x'',t'';x',t') = N_p(t'',t') \chi_p \left( -\frac{1}{h}
    \bar{S}(x'',t'';x',t')\right) ,   \eqno(14)
$$
where $N_p(t'',t')$ does not depend on end points $x''$ and  $x'$.
To calculate $N_p(t'',t')$ we use conditions (3) and (4).
Substituting ${\cal K}_p(x'',t'';x',t')$ into (4) we obtain
(for details, see \cite{branko2}):
$$
   N_p(t'',t') =  \left\vert\frac{1}{h} \frac{\partial^2\bar{S}}
   {\partial x''\partial x'}(x'',t'';x',t') \right\vert_p^{\frac{1}{2}}
   A_p(t'',t') ,\eqno(15)
$$
where $\vert A_p(t'',t')\vert_\infty =1$, \ \ ($   \vert \cdot\vert_p$  and $\vert\cdot\vert_\infty$ 
denote $p$-adic and absolute value, 
respectively). Replacing (15) in equation (3) we get conditions:
$$
   A_p(t'',t) A_p(t,t')\lambda_p (\alpha) = A_p(t'',t') ,    \eqno(16)
$$
$$
  \left\vert \frac{1}{h}\frac{\partial^2\bar{S}}{\partial x'' \partial x}(x'',t'';x,t)  
  \right\vert_p^{\frac{1}{2}}    \left\vert \frac{1}{h} \frac{\partial^2\bar{S}}{\partial x\partial x'}
  (x,t;x',t')\right\vert_p^{\frac{1}{2}}  \vert 2\alpha \vert_p^{-\frac{1}{2}}
  = \left\vert \frac{1}{h}  \frac{\partial^2\bar{S}}{\partial x'' \partial x'}
  (x'',t'';x',t')\right\vert_p^{\frac{1}{2}},          \eqno(17)
$$
where
$$
    \alpha = - \frac{1}{2h} \left[ \frac{\partial^2\bar{S}}{\partial x^2}(x'',t'';x,t)
    +\frac{\partial^2\bar{S}}{\partial x^2}(x,t;x',t')    \right] .                 \eqno(18)
$$    
Analysing the above formulae we obtain \cite{branko2}
$$
  A_p(t'',t') = \lambda_p\left( -\frac{1}{2h} \frac{\partial^2\bar{S}}{\partial x''\partial x'}
  (x'',t'';x',t')\right)  .    \eqno(19)
$$
For details of a  quite rigorous derivation of (19), see \cite{branko3}.

As the final result we have
$$
{\cal K}_p(x'',t'';x',t') = \lambda_p\left( -\frac{1}{2h} \frac{\partial^2\bar{S}}
{\partial x'' \partial x'} \right)   \left\vert \frac{1}{h} \frac{\partial^2\bar{S}}
{\partial x'' \partial x'}   \right\vert_p^{\frac{1}{2}}
\chi_p\left( -\frac{1}{h} \bar{S}(x'',t'';x',t')  \right)
\eqno(20)
$$
which is the $p$-adic Feynman path integral for quadratic
Lagrangians. The corresponding path integral of ordinary
quantum mechanics \cite{branko2} can be transformed into the same form as
(20), i.e. in such case index $p$ is replaced by index
$\infty$. This supports Volovich's conjecture \cite{volovich} that fundamental
physical laws should be invariant under interchange of number fields
$Q_p$ and $R$. 

\smallskip
{\bf Acknowledgement}.  Author wishes to thank the organizers of the
Bogolyubov Conference: Problems of Theoretical and Mathematical Physics, 
for invitation to participate in Moscow and Dubna parts of the Conference.

\end{document}